\documentclass[12pt,a4paper]{article}

\usepackage[a4paper,margin=2.5cm]{geometry}
\usepackage[T1]{fontenc}
\usepackage[utf8]{inputenc}
\usepackage{lmodern}
\usepackage{microtype}
\usepackage{amsmath,amssymb,bm}
\usepackage{graphicx}
\usepackage{booktabs}
\usepackage{hyperref}
\usepackage{cite}
\usepackage{setspace}
\hypersetup{hidelinks,pdftitle={Quantum-gravity-inspired effects in neutron stars: a critical map from microscopic assumptions to observables},pdfauthor={Randall Hell Vargas Pradinett},pdfsubject={Quantum-gravity phenomenology in neutron stars},pdfkeywords={quantum gravity, neutron stars, generalized uncertainty principle, noncommutative geometry, gravitational effective field theory}}

\title{Quantum-Gravity-Inspired Effects in Neutron Stars:\\
A Critical Map from Microscopic Assumptions to Observables}
\author{Randall Hell Vargas Pradinett\\
\small Institute for Theoretical Physics (IFT), S\~ao Paulo State University (UNESP),\\
\small S\~ao Paulo, SP, Brazil\\
\small \texttt{randall.vargas@unesp.br}}
\date{}

\begin{document}
\maketitle

\begin{abstract}
Claims that neutron stars constrain quantum gravity often compare models whose new physics enters at different stages of the stellar calculation. This article provides a critical map of three representative routes: generalized-uncertainty-principle deformations of microscopic thermodynamics, noncommutative-inspired modifications of the effective matter source, and higher-curvature or scale-dependent modifications of gravitational dynamics. For each route we identify the quantity that is changed, the dimensionless parameter controlling the correction, and the consistency conditions required before masses, radii, or tidal deformabilities can be compared with observations. A deformed phase-space measure must be implemented in a thermodynamically consistent interacting equation of state; a smeared density must be completed by a conserved stress-energy tensor, generally with anisotropic pressure; and a gravitational effective field theory must remain within its derivative expansion and be applied consistently to both the equilibrium background and tidal perturbations. Simple scale estimates show that coefficients of natural size suppressed by the Planck scale are negligible in neutron stars. Observable effects consequently constrain model-dependent effective scales or enhanced couplings, not microscopic quantum gravity directly. The resulting framework is intended to separate robust phenomenological bounds from effects produced by inconsistent implementations or by degeneracies with the high-density equation of state.
\end{abstract}

\noindent\textbf{Keywords:} quantum gravity; neutron stars; generalized uncertainty principle; noncommutative geometry; effective field theory; equation of state; modified gravity; multimessenger astronomy

\section{Introduction}
\label{sec:introduction}

General Relativity (GR) provides an exceptionally successful description
of gravity on macroscopic scales, while quantum mechanics and quantum
field theory describe matter and the nongravitational interactions with
remarkable accuracy. Despite their individual success, these frameworks
are not expected to remain independent at arbitrarily high energies. A
consistent description of gravity in the quantum regime remains one of
the central open problems in theoretical physics
~\cite{AmelinoCamelia2013,Hossenfelder2013}.

The characteristic scale at which quantum-gravitational effects are
usually expected to become important follows from the constants
$\hbar$, $G$, and $c$. The corresponding Planck length, mass, and energy
are
\begin{align}
\ell_{\rm Pl}
&=
\sqrt{\frac{\hbar G}{c^3}},
&
M_{\rm Pl}
&=
\sqrt{\frac{\hbar c}{G}},
\label{eq:planck_scales}
\\
E_{\rm Pl}
&=
M_{\rm Pl}c^2
=
\sqrt{\frac{\hbar c^5}{G}}.
\label{eq:planck_energy}
\end{align}
Here $M_{\rm Pl}$ denotes the unreduced Planck mass. Numerically,
\begin{align}
\ell_{\rm Pl}
&\simeq
1.6\times10^{-35}\ {\rm m},
&
M_{\rm Pl}
&\simeq
2.2\times10^{-8}\ {\rm kg},
\nonumber\\
E_{\rm Pl}
&\simeq
1.2\times10^{19}\ {\rm GeV}.
\label{eq:planck_values}
\end{align}
These scales are far beyond those accessible to present experiments.
This separation has motivated a broad program of quantum-gravity
phenomenology: rather than attempting to probe the microscopic quantum
structure of spacetime directly, one asks whether effective remnants of
the underlying theory could survive at much lower energies and leave
observable signatures~\cite{AmelinoCamelia2013,Hossenfelder2013}.

Compact astrophysical objects provide a natural setting in which to
explore this possibility. Neutron stars are particularly interesting
because they combine several physical regimes that are difficult to
reproduce simultaneously in terrestrial experiments. Their interiors
contain strongly interacting matter at densities of several times
nuclear saturation density, while their global structure is governed by
strong gravitational fields and therefore requires a relativistic
description. At the same time, neutron stars are increasingly accessible
to observation. Pulsar timing provides precise mass measurements,
X-ray pulse-profile modeling constrains stellar masses and radii, and
gravitational waves from binary neutron-star mergers probe their tidal
response. Neutron stars have consequently become important laboratories
for dense nuclear matter, strong-field gravity, and possible departures
from their standard descriptions
~\cite{LattimerPrakash2004,Oertel2017,Baym2018,Burgio2021}.

This possibility must nevertheless be interpreted with care. Even the
densest regions of neutron stars remain enormously separated from the
Planck scale. Observable deviations should therefore not be interpreted
as direct manifestations of the microscopic quantum structure of
spacetime. The relevant question is instead whether a given
quantum-gravity-inspired scenario generates an effective correction that
remains sufficiently enhanced at neutron-star energies, densities, or
curvature scales to influence macroscopic stellar properties. This point
is especially relevant because many phenomenological constructions
introduce effective scales or couplings that need not coincide with
their naive Planck-scale values.

A useful way of organizing these effects is according to where the new
physics enters the stellar-structure problem. Quantum-gravity-inspired
corrections can act on the microscopic matter sector, on the effective
description of the matter source, or directly on the gravitational
dynamics. These possibilities are conceptually distinct, although their
observable consequences can overlap.

If the correction acts on the microscopic matter sector, modifications
of kinematics, phase space, or statistical mechanics change the
thermodynamic quantities entering the equation of state (EOS),
\begin{equation}
P=P(\varepsilon),
\label{eq:eos_intro}
\end{equation}
and therefore modify the stellar configuration even when the
gravitational field equations retain their standard GR form. In a
second class of models, the localization or stress-energy distribution
of the source itself is modified, potentially leading to extended or
anisotropic matter configurations. Finally, modifications of the
gravitational sector alter the field equations and hence the conditions
for hydrostatic equilibrium, even for an otherwise unchanged matter
EOS. These different routes from microscopic assumptions to
macroscopic observables are summarized schematically in
Fig.~\ref{fig:qg_chain}. The separation between matter, source, and
gravitational effects provides the organizing principle adopted
throughout this article.

An important complication is that these mechanisms do not necessarily
produce unique observational signatures. A modification of the EOS can,
for example, shift the mass--radius relation in a way that resembles the
effect of modified gravitational dynamics. Similar degeneracies can
arise in the tidal response. A deviation in a single observable would
therefore generally be insufficient to identify its microscopic origin.
Meaningful constraints require several observables to be considered
together with the uncertainties associated with the high-density EOS.

We examine three representative classes of quantum-gravity-inspired
phenomenology. We first consider generalized uncertainty principles
(GUPs), focusing on realizations in which the deformation modifies the
phase-space measure and, consequently, the thermodynamics of degenerate
matter. We then discuss noncommutative-inspired descriptions, where a
fundamental limitation on localization can be represented through an
effective smearing of the matter source and may lead to modified or
anisotropic stress-energy distributions. Finally, we turn to the
gravitational sector and discuss higher-curvature corrections from the
perspective of gravitational effective field theory (EFT), together
with phenomenological scenarios involving scale-dependent gravitational
couplings.

This is a focused critical perspective rather than an exhaustive
catalogue. Its contribution is twofold. First, it places three
frequently conflated mechanisms in one chain from microscopic assumption
to observable. Second, it formulates checks that any claimed
neutron-star bound must pass: thermodynamic consistency, stress-energy
conservation, control of the effective expansion, and a joint treatment
of dense-matter uncertainty. A visible numerical change obtained by
choosing a large effective parameter is not, by itself, evidence for an
amplified Planck-scale correction; it constrains that effective model.

\section{Neutron stars as laboratories for quantum-gravity phenomenology}
\label{sec:NS}

Neutron stars provide a unique environment in which dense nuclear matter
and strong gravitational fields coexist. They are the compact remnants
of massive stars and typically contain masses of order
$1$--$2\,M_\odot$ within radii of approximately
$10$--$15~\mathrm{km}$. Their interiors can reach baryon densities of
several times the nuclear saturation density,
\begin{equation}
n_0 \simeq 0.16~\mathrm{fm}^{-3},
\label{eq:nuclear_saturation}
\end{equation}
placing the stellar core in a regime where the composition and
properties of strongly interacting matter remain uncertain
\cite{LattimerPrakash2004,Oertel2017,Baym2018,Burgio2021}. This
combination of high density, strong gravity, and direct astrophysical
access makes neutron stars particularly useful for testing extensions
of both dense-matter physics and gravitational theory.

A convenient measure of the strength of the gravitational field is the
stellar compactness,
\begin{equation}
C=\frac{GM}{Rc^2}.
\label{eq:compactness}
\end{equation}
For a typical neutron star, $C\sim0.1$--$0.3$, so relativistic effects
are essential for determining the stellar structure. In GR, a
nonrotating neutron star can be described, to a good approximation, as
a static and spherically symmetric configuration. The spacetime metric
is then written as
\begin{equation}
ds^2
=
-e^{2\Phi(r)}c^2dt^2
+
\left(
1-\frac{2Gm(r)}{rc^2}
\right)^{-1}dr^2
+
r^2d\Omega^2,
\label{eq:ns_metric}
\end{equation}
where $m(r)$ is the gravitational mass enclosed within radius $r$ and
$\Phi(r)$ determines the temporal component of the metric.

For an isotropic perfect fluid, Einstein's equations reduce to the
Tolman--Oppenheimer--Volkoff (TOV) equations
\cite{Tolman1939,OppenheimerVolkoff1939},
\begin{align}
\frac{dm}{dr}
&=
4\pi r^2\frac{\epsilon(r)}{c^2},
\label{eq:tov_mass}
\\
\frac{dP}{dr}
&=
-
\frac{G}{c^2}
\frac{
\left[\epsilon(r)+P(r)\right]
\left[
m(r)+\dfrac{4\pi r^3P(r)}{c^2}
\right]
}{
r^2
\left[
1-\dfrac{2Gm(r)}{rc^2}
\right]
},
\label{eq:tov_pressure}
\end{align}
where $\epsilon(r)$ and $P(r)$ denote the local energy density and
pressure. The system is closed by specifying an equation of state,
\begin{equation}
P=P(\epsilon).
\label{eq:ns_eos}
\end{equation}
Starting from a central pressure $P_c$, the equations are integrated
outward with $m(0)=0$ until the pressure vanishes at $r=R$. The total
gravitational mass is then $M=m(R)$. Repeating this procedure for
different central pressures generates the mass--radius sequence
associated with a given EOS \cite{LattimerPrakash2004,Oertel2017}.

This construction also makes clear where physics beyond the standard
description can enter. A modification of the microscopic matter sector
changes the relation $P(\epsilon)$ while leaving the TOV equations
unchanged. A modification of gravity instead changes the stellar
structure equations themselves. Other scenarios may modify the
effective stress-energy tensor and therefore alter the source entering
the gravitational equations. Although these possibilities are
theoretically distinct, they can lead to similar changes in macroscopic
observables. This is one of the main difficulties in using neutron stars
to constrain new physics.

The connection with observations has improved substantially in recent
years. Precise radio timing has established neutron stars with masses
around two solar masses, placing a robust lower bound on the maximum
mass supported by any viable stellar model
\cite{Demorest2010,Antoniadis2013,Cromartie2020}. X-ray pulse-profile
modeling provides simultaneous information on masses and radii
\cite{Miller2019,Riley2019,Miller2021,Riley2021}, while
gravitational-wave observations of binary neutron-star mergers probe
the tidal response of the stars \cite{Abbott2017,Abbott2018}. These
observables test different aspects of the stellar structure and become
considerably more informative when analyzed together.

The usefulness of neutron stars for quantum-gravity phenomenology
should, however, be understood in the context of the enormous hierarchy
between stellar and Planck scales. Typical densities in the inner core
are of order
\begin{equation}
\rho_{\rm NS}\sim10^{17}-10^{18}\ {\rm kg\,m^{-3}},
\label{eq:ns_density_scale}
\end{equation}
whereas the Planck density is
\begin{equation}
\rho_{\rm Pl}
=
\frac{c^5}{\hbar G^2}
\simeq
5.2\times10^{96}\ {\rm kg\,m^{-3}}.
\label{eq:planck_density}
\end{equation}
The ratio is therefore roughly
\begin{equation}
\frac{\rho_{\rm NS}}{\rho_{\rm Pl}}
\sim 10^{-80}-10^{-79}.
\end{equation}
A similarly large separation occurs at the level of characteristic
curvature scales. Consequently, neutron stars should not be interpreted as
objects approaching the microscopic Planck regime
\cite{AmelinoCamelia2013,Hossenfelder2013}.

Their role is instead to test whether a proposed theory or
phenomenological model contains corrections that remain relevant far
below the Planck scale. Such corrections may alter the microscopic
thermodynamics of dense matter, the effective stress-energy tensor, or
the gravitational dynamics. If sufficiently large, these changes
propagate through the stellar structure problem and can modify
quantities such as the maximum mass, radius, moment of inertia, tidal
deformability, or oscillation spectrum.

This last point is important when interpreting phenomenological bounds.
If a correction is naturally suppressed by powers of
$\ell_{\rm Pl}/L$, $E/E_{\rm Pl}$, or the corresponding curvature
ratio, its effect in a neutron star will generally be extremely small.
An observable modification therefore requires either an enhancement
mechanism, a new effective scale well separated from the Planck scale,
or a coupling substantially larger than its naive Planck-suppressed
value. Constraints obtained from neutron stars should accordingly be
understood primarily as bounds on the effective parameters of the model
being tested, rather than as direct measurements of microscopic
quantum-gravity physics.

There is a second difficulty. Uncertainties in the high-density EOS can
produce changes in neutron-star observables that resemble those caused
by modified gravity or other new physics. A change in the radius or
maximum mass, for example, is not by itself sufficient to determine
whether the underlying cause lies in the matter sector or in the
gravitational dynamics. Tidal observables introduce additional
information, but they are also sensitive to the EOS. Reliable
constraints therefore require several complementary observables and a
consistent treatment of both dense-matter and gravitational
uncertainties.

These considerations define the perspective adopted in the following
sections. Rather than asking whether neutron stars reach the quantum
gravity scale, we ask a more restricted and phenomenologically useful
question: how do specific quantum-gravity-inspired corrections enter
the stellar problem, what effective scales control their magnitude,
and under what conditions could their consequences become observable?

\section{How quantum gravity enters neutron-star calculations}
\label{sec:framework}

Quantum-gravity-inspired physics does not translate directly into a
neutron-star mass, radius, or tidal deformability. Between a microscopic
modification and an astrophysical observable lies the description of
dense matter, the gravitational source, and the equations governing
stellar equilibrium. It is therefore useful to organize phenomenological
models according to where the new physics enters the neutron-star
calculation.

\begin{figure}[t]
\centering
\includegraphics[width=0.95\columnwidth]
{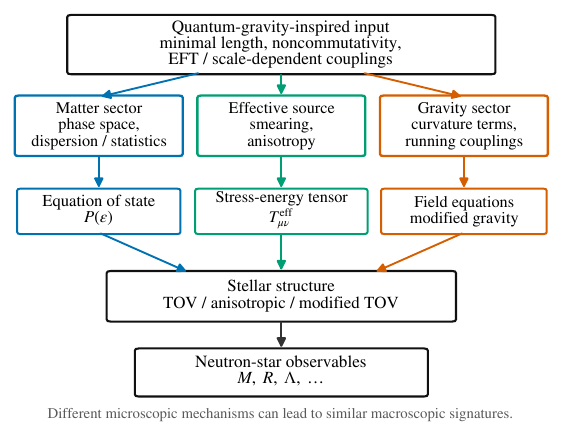}
\caption{
Schematic connection between quantum-gravity-inspired physics and
neutron-star observables. New physics may enter through the microscopic
matter sector, modifying the equation of state; through an effective
description of the matter source, modifying the stress-energy tensor;
or through the gravitational sector, modifying the field equations.
Each route changes the stellar-structure problem and can ultimately
affect observables such as the mass $M$, radius $R$, and tidal
deformability $\Lambda$. The resulting signatures are therefore
indirect and generally model dependent.
}
\label{fig:qg_chain}
\end{figure}

As summarized in Fig.~\ref{fig:qg_chain}, three conceptually distinct
routes can be identified. A correction may modify the microscopic matter
sector and hence the equation of state (EOS); it may alter the effective
stress-energy tensor that acts as the gravitational source; or it may
change the gravitational field equations themselves. These possibilities
are not mutually exclusive, but separating them provides a useful
framework for comparing models with otherwise very different microscopic
motivations.

\subsection{Matter-sector modifications}
\label{subsec:matter_route}

The first route leaves the gravitational dynamics unchanged but modifies
the microscopic description of matter. Changes in the phase-space
measure, dispersion relation, statistical mechanics, or interactions
alter the thermodynamic quantities from which the EOS is constructed.
Schematically,
\begin{equation}
P(\epsilon)
\longrightarrow
P_{\rm QG}(\epsilon;\alpha),
\label{eq:qg_eos}
\end{equation}
where $\alpha$ denotes the parameter, or set of parameters, controlling
the deformation.

Generalized uncertainty principles provide a representative example.
Depending on the realization, a modified commutation relation changes
the density of accessible momentum states and therefore the
thermodynamics of degenerate matter
~\cite{Maggiore1993,Kempf1995,AliDasVagenas2009}. Once the modified
pressure and energy density have been obtained consistently, the
resulting EOS can be used in the standard GR stellar-structure
equations.

This route is conceptually important because it shows that a change in
neutron-star structure does not necessarily imply a modification of
gravity. New microscopic physics can shift the mass--radius relation or
tidal response entirely through its effect on the EOS.

\subsection{Effective-source modifications}
\label{subsec:source_route}

A second possibility is that the new physics is represented through an
effective gravitational source. In this case, the relevant modification
can be written schematically as
\begin{equation}
T_{\mu\nu}
\longrightarrow
T_{\mu\nu}^{\rm eff}(\alpha).
\label{eq:qg_source}
\end{equation}
The gravitational field equations may retain their Einstein form, but
the source entering them is no longer described by the conventional
perfect-fluid stress-energy tensor.

This situation occurs naturally in phenomenological models in which a
fundamental localization scale is represented by replacing a sharply
localized source with an extended distribution. Depending on the
construction, the effective source may also develop anisotropic stresses,
so that the radial and tangential pressures are no longer equal. The
stellar equilibrium equations must then be modified consistently with
the form of $T_{\mu\nu}^{\rm eff}$.

The noncommutative-inspired models discussed in
Sec.~\ref{sec:noncomm} illustrate this route. Their role in the present
article is not to provide a unique prediction of noncommutative field
theory, but to show how a microscopic limitation on localization can be
encoded phenomenologically in the effective source of a compact star.

\subsection{Modifications of the gravitational sector}
\label{subsec:gravity_route}

The third route changes the gravitational dynamics directly. The matter
EOS may remain unchanged while the Einstein equations acquire additional
terms,
\begin{equation}
G_{\mu\nu}
+
\Delta_{\mu\nu}[g;\alpha]
=
\frac{8\pi G}{c^4}T_{\mu\nu},
\label{eq:qg_gravity}
\end{equation}
where $\Delta_{\mu\nu}$ represents the correction generated by the
effective gravitational theory. Higher-curvature operators and
scale-dependent gravitational couplings are representative examples
~\cite{Donoghue1994,Burgess2004,
SotiriouFaraoni2010,DeFeliceTsujikawa2010}.

In this case, the stellar-structure problem itself changes. The
hydrostatic-equilibrium equations must be derived from the modified
field equations and their associated conservation laws rather than
obtained by simply inserting a new parameter into the standard TOV
equation. If tidal properties are calculated, the corresponding
perturbation equations must likewise be treated within the same
gravitational framework.

The three routes shown in Fig.~\ref{fig:qg_chain} therefore modify
different ingredients of the calculation,
\begin{equation}
\begin{aligned}
\bigl\{
P(\epsilon),\,
T_{\mu\nu}^{\rm eff},\,
\text{field equations}
\bigr\}
&\longrightarrow
\text{stellar structure}
\\
&\longrightarrow
\bigl\{
M,\,
R,\,
\Lambda,\ldots
\bigr\}.
\end{aligned}
\label{eq:qg_pipeline}
\end{equation}
This separation will be used throughout the remainder of the article.
It also makes clear why astrophysical constraints are inherently
model dependent: physically different microscopic mechanisms can
produce similar changes in the same macroscopic observables.

\begin{table}[ht]
\centering
\caption{Three entry points for quantum-gravity-inspired corrections.}
\label{tab:three_routes}
\small
\begin{tabular}{p{0.17\textwidth}p{0.19\textwidth}p{0.17\textwidth}p{0.34\textwidth}}
\toprule
Route & Modified object & Control parameter & Minimum consistency requirement \\
\midrule
GUP matter sector & phase space, dispersion, or statistics & $\beta p_F^2$ & derive $n$, $\varepsilon$, and $P$ from one potential and solve the interacting composition self-consistently \\
NC-inspired source & $T_{\mu\nu}^{\rm eff}$ and localization & $\theta/L^2$ & define the kernel, close the pressure sector, and impose $\nabla_\mu T^{\mu\nu}_{\rm eff}=0$ \\
Gravitational EFT & field equations and perturbations & $\mathcal{R}/M^2$ & control the expansion, remove spurious branches, and use the same theory for the background and tides \\
\bottomrule
\end{tabular}
\end{table}

\subsection{The equation of state as the microscopic input}
\label{subsec:eos_bridge}

For corrections that enter through the matter sector, the EOS provides
the bridge between microscopic physics and stellar structure
~\cite{LattimerPrakash2004,Oertel2017,Baym2018,Burgio2021}.
Although observations constrain global quantities such as masses and
radii, these quantities depend on the pressure generated by matter over
a broad range of densities. Any microscopic deformation that changes
the pressure, energy density, or composition can therefore propagate
to observable neutron-star properties.

For matter in thermodynamic equilibrium, it is convenient to introduce
the grand-potential density
\begin{equation}
\omega(T,\mu)
=
\frac{\Omega}{V},
\label{eq:grand_density}
\end{equation}
from which
\begin{equation}
P=-\omega,
\qquad
n=
-\left(
\frac{\partial\omega}{\partial\mu}
\right)_T,
\label{eq:thermo_pressure_density}
\end{equation}
and
\begin{equation}
\epsilon
=
\omega+Ts+\mu n.
\label{eq:thermo_energy}
\end{equation}
Here $n$ and $s$ are the particle-number and entropy densities,
respectively. In the zero-temperature limit relevant for cold
neutron-star matter,
\begin{equation}
\epsilon
=
\omega+\mu n.
\label{eq:thermo_energy_t0}
\end{equation}
Eliminating the microscopic variables then yields the relation
$P=P(\epsilon)$ required by the stellar-structure equations.

A relativistic degenerate Fermi gas provides a transparent illustration.
At zero temperature,
\begin{equation}
\omega
=
-g
\int
\frac{d^3p}{(2\pi\hbar)^3}
\left[
\mu-E(p)
\right]
\Theta(p_F-p),
\label{eq:fermi_grand_potential}
\end{equation}
with
\begin{equation}
E(p)
=
\sqrt{p^2c^2+m^2c^4},
\label{eq:relativistic_dispersion}
\end{equation}
where $g$ is the degeneracy factor and $p_F$ is the Fermi momentum.
For the ideal gas at $T=0$, $\mu=E(p_F)$.

Equation~\eqref{eq:fermi_grand_potential} makes explicit the microscopic
ingredients that can be modified. A deformation of the phase-space
measure changes the weight assigned to momentum states, a modified
dispersion relation changes their single-particle energies, and new
interactions contribute directly to the thermodynamic potential. These
changes need not lead to the same EOS even when they arise from related
ideas at the microscopic level.

Realistic neutron-star matter is considerably more complicated than an
ideal one-component Fermi gas. Strong interactions, the particle
composition, charge neutrality, and beta equilibrium all contribute to
the EOS. Equation~\eqref{eq:fermi_grand_potential} should therefore be
viewed only as a simple baseline for identifying where a microscopic
deformation enters. In a realistic calculation, the same deformation
must be incorporated consistently into all thermodynamic quantities that
it affects.

The remainder of the article follows the three routes introduced above.
Generalized uncertainty principles illustrate modifications of the
matter sector; noncommutative-inspired constructions illustrate changes
in the effective source; and the effective-field-theory treatment of
gravity illustrates modifications of the gravitational dynamics. This
organization makes it possible to compare their neutron-star
consequences within a common framework while keeping their physical
origins distinct.

\section{Generalized uncertainty principle}
\label{sec:gup}

A recurring idea in quantum-gravity phenomenology is that the notion of
arbitrarily short distances may cease to be physically meaningful.
Arguments based on quantum gravity, string-inspired models, and
thought experiments combining quantum mechanics with gravitational
collapse have motivated the possibility of an effective minimum
measurable length~\cite{Garay1995,Hossenfelder2013,Maggiore1993,Kempf1995}.
Generalized uncertainty principles (GUPs) provide a convenient
phenomenological framework for describing such a possibility. They
should not, in general, be regarded as a complete microscopic theory of
quantum spacetime, but rather as effective deformations through which
possible short-distance effects can be studied at lower energies.

In ordinary quantum mechanics, the canonical algebra
\begin{equation}
[x_i,p_j]
=
i\hbar\delta_{ij}
\label{eq:canonical_commutator}
\end{equation}
leads to the Heisenberg uncertainty relation
\begin{equation}
\Delta x\,\Delta p
\geq
\frac{\hbar}{2}.
\label{eq:heisenberg}
\end{equation}
Within this framework, the uncertainty in position can in principle be
made arbitrarily small by increasing the uncertainty in momentum.
Gravity changes the physical interpretation of this limiting procedure:
localizing a state within an increasingly small region requires an
increasing concentration of energy, which must eventually contribute
non-negligibly to the surrounding spacetime geometry
~\cite{Garay1995,Scardigli1999}. This observation motivates
deformations of the canonical algebra in which momentum-dependent
corrections become important at sufficiently high momentum.

A commonly used schematic realization is
\begin{equation}
[x_i,p_j]
=
i\hbar\delta_{ij}
\left(1+\beta p^2+\cdots\right),
\label{eq:gupcommutator}
\end{equation}
where $\beta$ controls the strength of the deformation. This expression
is deliberately simplified. In more general multidimensional
constructions, the commutator can contain additional tensor structures,
including terms proportional to $p_i p_j$, and may involve more than
one deformation parameter~\cite{Kempf1995,Hossenfelder2013}.
Consequently, neither the precise form of the algebra nor its
phenomenological consequences are universal among GUP models.

At the level of the uncertainty relation, Eq.~\eqref{eq:gupcommutator}
suggests a modification of the form
\begin{equation}
\Delta x\,\Delta p
\gtrsim
\frac{\hbar}{2}
\left[
1+\beta(\Delta p)^2+\cdots
\right].
\label{eq:guprelation}
\end{equation}
Minimizing the right-hand side with respect to the momentum uncertainty
gives a characteristic minimum position uncertainty of order
\begin{equation}
(\Delta x)_{\rm min}
\sim
\hbar\sqrt{\beta}.
\label{eq:gup_min_length}
\end{equation}
It is customary to parametrize the deformation as
\begin{equation}
\beta
=
\frac{\beta_0}{(M_{\rm Pl}c)^2},
\label{eq:beta_definition}
\end{equation}
where $\beta_0$ is dimensionless and $M_{\rm Pl}$ is the unreduced
Planck mass. The parameter $\beta$ therefore has dimensions of inverse
momentum squared, and the relevant expansion parameter is $\beta p^2$.

If $\beta_0$ is of order unity, the deformation becomes appreciable only
for momenta approaching the Planck scale. For characteristic
neutron-star momenta, this produces an enormous suppression. Taking
$p\sim 1~{\rm GeV}/c$ only as an order-of-magnitude scale,
\begin{equation}
\beta p^2
=
\beta_0
\left(
\frac{pc}{M_{\rm Pl}c^2}
\right)^2
\sim
10^{-38}\,\beta_0.
\label{eq:gup_ns_scale}
\end{equation}
Thus, for $\beta_0\sim\mathcal{O}(1)$, a Planck-suppressed GUP
correction is effectively invisible in neutron-star matter. Observable
stellar effects require either a very large effective value of
$\beta_0$ or a deformation associated with a scale substantially below
the Planck scale. Astrophysical constraints on $\beta$ should therefore
be interpreted primarily as bounds on an effective GUP realization,
rather than as direct measurements of the Planck length
~\cite{Hossenfelder2013,WangYang2012}. An idealized GUP Fermi-gas
analysis obtained $\beta_0\lesssim10^{37}$ from neutron-star masses,
while emphasizing that nuclear interactions can produce comparable
macroscopic changes \cite{WangYang2012}. This limit is conditional on
the adopted matter model; it is not evidence that a parameter of natural
size has become observable.

\subsection{Modified phase-space measure}
\label{subsec:gup_phase_space}

One possible consequence of a deformed quantum algebra is a modification
of the density of available states in momentum space. This is
particularly relevant for degenerate matter because its thermodynamic
properties are determined by integrals over occupied momentum states.
A deformation of the phase-space measure can therefore modify the
equation of state even when the single-particle dispersion relation and
the gravitational field equations are left unchanged.

The relation between a GUP algebra and the corresponding phase-space
measure depends on the representation chosen for the deformed
commutation relations. There is no unique measure associated with all
GUP models. To make the thermodynamic consequences explicit, we adopt
the representative isotropic prescription
\begin{equation}
\frac{d^3p}{(2\pi\hbar)^3}
\longrightarrow
\frac{d^3p}
{(2\pi\hbar)^3(1+\beta p^2)^3}.
\label{eq:gupmeasure}
\end{equation}
The additional factor suppresses the contribution of high-momentum
states. Equation~\eqref{eq:gupmeasure} should therefore be understood
as a particular phenomenological realization of GUP-modified phase
space, rather than as a model-independent prediction of the generalized
uncertainty principle~\cite{Kempf1995,Hossenfelder2013}.
For this algebra, the cubic power follows from the invariant
three-dimensional phase-space volume obtained from the corresponding
classical Liouville theorem \cite{Chang2002}; changing the algebra or
the number of dimensions generally changes the Jacobian.

\begin{figure}[t]
\centering
\includegraphics[width=\columnwidth]
{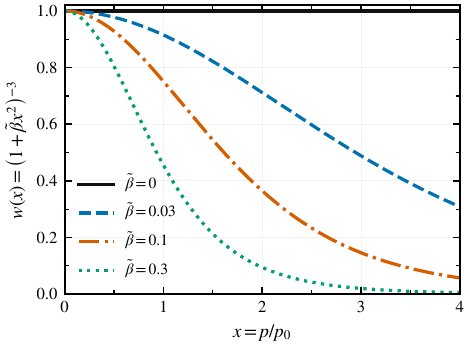}
\caption{Dimensionless phase-space weight
$w(x)=(1+\widetilde{\beta}x^2)^{-3}$ for representative values of the
deformation parameter $\widetilde{\beta}=\beta p_0^2$, with
$x=p/p_0$. The canonical phase-space measure is recovered for
$\widetilde{\beta}=0$. Positive values of $\widetilde{\beta}$
progressively suppress high-momentum states. The curves illustrate the
specific phase-space realization adopted in Eq.~\eqref{eq:gupmeasure}
and should not be interpreted as a universal prediction of GUP
models.}
\label{fig:gup_phase_space}
\end{figure}

To display the deformation independently of a particular momentum
unit, we introduce a reference scale $p_0$ and define
\begin{equation}
x\equiv\frac{p}{p_0},
\qquad
\widetilde{\beta}\equiv\beta p_0^2.
\label{eq:gup_dimensionless_variables}
\end{equation}
The phase-space weight then becomes
\begin{equation}
w(x)
=
\left(1+\widetilde{\beta}x^2\right)^{-3}.
\label{eq:gup_weight_dimensionless}
\end{equation}
Figure~\ref{fig:gup_phase_space} shows this weight for several
representative values of $\widetilde{\beta}$. The undeformed limit is
recovered for $\widetilde{\beta}=0$, whereas increasing the deformation
progressively suppresses states at large momentum.

The thermodynamic consequences can be seen most transparently for a
relativistic Fermi gas at zero temperature. In the undeformed case, the
number density is
\begin{equation}
n
=
\frac{g}{2\pi^2\hbar^3}
\int_0^{p_F}p^2\,dp
=
\frac{g p_F^3}{6\pi^2\hbar^3},
\label{eq:number_density_standard}
\end{equation}
where $g$ is the degeneracy factor and $p_F$ is the Fermi momentum. The
corresponding energy density and pressure are
\begin{align}
\epsilon
&=
\frac{g}{2\pi^2\hbar^3}
\int_0^{p_F}
p^2E(p)\,dp,
\label{eq:energy_density_standard}
\\
P
&=
\frac{g}{6\pi^2\hbar^3}
\int_0^{p_F}
\frac{p^4c^2}{E(p)}\,dp,
\label{eq:pressure_standard}
\end{align}
with
\begin{equation}
E(p)
=
\sqrt{p^2c^2+m^2c^4}.
\label{eq:gup_relativistic_energy}
\end{equation}

Using the modified measure in Eq.~\eqref{eq:gupmeasure}, the
particle-number density becomes
\begin{equation}
n_{\rm GUP}
=
\frac{g}{2\pi^2\hbar^3}
\int_0^{p_F}
\frac{p^2\,dp}
{(1+\beta p^2)^3},
\label{eq:gup_number_density}
\end{equation}
while the energy density is
\begin{equation}
\epsilon_{\rm GUP}
=
\frac{g}{2\pi^2\hbar^3}
\int_0^{p_F}
\frac{p^2E(p)}
{(1+\beta p^2)^3}\,dp.
\label{eq:gup_energy_density}
\end{equation}

The pressure is most conveniently obtained from the same thermodynamic
potential rather than introduced through an independent deformation of
the kinetic-pressure integral. At zero temperature, the grand-potential
density for the noninteracting gas is
\begin{equation}
\omega_{\rm GUP}
=
-\frac{g}{2\pi^2\hbar^3}
\int_0^{p_F}
\frac{p^2[\mu-E(p)]}
{(1+\beta p^2)^3}\,dp,
\label{eq:gup_grand_potential}
\end{equation}
where
\begin{equation}
\mu=E(p_F).
\label{eq:gup_chemical_potential}
\end{equation}
The pressure follows from
\begin{equation}
P_{\rm GUP}
=
-\omega_{\rm GUP},
\label{eq:gup_pressure}
\end{equation}
and the number density is obtained consistently from
\begin{equation}
n_{\rm GUP}
=
-
\left(
\frac{\partial\omega_{\rm GUP}}{\partial\mu}
\right)_{T=0}.
\label{eq:gup_density_thermo}
\end{equation}
These expressions recover the ordinary relativistic Fermi-gas
thermodynamics in the limit $\beta\rightarrow0$.

The deformation changes the relation among $n$, $\epsilon$, and $P$ and
therefore modifies the EOS. In the class of models considered here,
this is the point at which the GUP enters the neutron-star calculation:
the microscopic thermodynamics is modified, while the Einstein field
equations and the standard TOV equations remain unchanged.

\subsection{Perturbative limit and characteristic size of the correction}
\label{subsec:gup_perturbative}

The magnitude of the effect is particularly transparent in the
perturbative regime
\begin{equation}
\beta p_F^2\ll1.
\label{eq:gup_perturbative_condition}
\end{equation}
Expanding the phase-space weight gives
\begin{equation}
(1+\beta p^2)^{-3}
=
1-3\beta p^2
+6\beta^2p^4
+\mathcal{O}(\beta^3p^6).
\label{eq:gup_weight_expansion}
\end{equation}
Keeping the leading correction in Eq.~\eqref{eq:gup_number_density}
yields
\begin{equation}
n_{\rm GUP}
\simeq
\frac{g}{2\pi^2\hbar^3}
\left(
\frac{p_F^3}{3}
-
\frac{3\beta p_F^5}{5}
\right).
\label{eq:gup_number_density_expansion}
\end{equation}
Relative to the undeformed density evaluated at the same Fermi
momentum,
\begin{equation}
n
=
\frac{g p_F^3}{6\pi^2\hbar^3},
\end{equation}
the leading fractional correction is
\begin{equation}
\frac{\delta n}{n}
\equiv
\frac{n_{\rm GUP}-n}{n}
\simeq
-\frac{9}{5}\beta p_F^2.
\label{eq:gup_relative_correction}
\end{equation}
The sign reflects the suppression of the available phase-space volume
in the realization adopted here, while the magnitude is controlled by
the dimensionless combination $\beta p_F^2$.

Equation~\eqref{eq:gup_relative_correction} refers specifically to a
comparison performed at fixed $p_F$. This distinction matters when
constructing an EOS. If the physical baryon density $n_B$ is specified
instead, the Fermi momentum is no longer the same in the deformed and
undeformed systems. It must be determined self-consistently from
\begin{equation}
n_B
=
\frac{g}{2\pi^2\hbar^3}
\int_0^{p_F}
\frac{p^2\,dp}
{(1+\beta p^2)^3}.
\label{eq:gup_fixed_density}
\end{equation}
The corresponding $\epsilon_{\rm GUP}$ and $P_{\rm GUP}$ must then be
evaluated using this value of $p_F$. A fixed-$p_F$ expansion is useful
for identifying the scale and sign of the microscopic correction, but
it should not be confused with a self-consistent comparison of
equations of state at fixed physical density.

\subsection{Connection with realistic dense-matter models}
\label{subsec:gup_realistic_matter}

The free Fermi-gas calculation isolates the effect of the modified
phase-space measure, but it is not a realistic model of a neutron-star
core. Dense stellar matter involves strong interactions, several
particle species, beta equilibrium, charge neutrality, and many-body
correlations, all of which contribute to the EOS
~\cite{Oertel2017,Baym2018,Burgio2021}.

A realistic implementation must therefore introduce the deformation
consistently within the chosen microscopic model. Schematically, the
grand-potential density may take the form
\begin{equation}
\omega_{\rm GUP}
=
-g
\int
\frac{d^3p}
{(2\pi\hbar)^3(1+\beta p^2)^3}
\,\mathcal{F}(p,\mu,M^\ast,\ldots)
+
\omega_{\rm int},
\label{eq:gup_grand_potential_interacting}
\end{equation}
where $\mathcal{F}$ represents the appropriate single-particle
contribution and $\omega_{\rm int}$ contains the interaction terms.
Equation~\eqref{eq:gup_grand_potential_interacting} is schematic by
construction. Depending on the microscopic theory and on the chosen
GUP realization, the deformation may also enter effective masses, mean
fields, particle fractions, chemical-equilibrium conditions, and the
self-consistency equations that determine the state of the matter.

This dependence on the underlying dense-matter model is important when
astrophysical bounds are interpreted. A GUP-induced change in the EOS
can be partially degenerate with uncertainties in nuclear interactions,
many-body correlations, the particle content of the core, or possible
high-density phase transitions. A bound on $\beta$ obtained from a
mass, radius, or tidal-deformability measurement is therefore also
conditional on the microscopic EOS used in the calculation.

For this reason, neutron-star constraints on GUP-inspired physics are
most meaningful when three requirements are satisfied simultaneously:
the deformation is implemented thermodynamically consistently, the
underlying dense-matter model is realistic, and several complementary
astrophysical observables are considered. Within such a framework,
neutron stars can constrain the effective scale associated with the
deformation, even though the natural Planck-suppressed correction itself
is far too small to produce an observable stellar effect
~\cite{Oertel2017,Baym2018,Burgio2021,Abbott2017,Abbott2018}.

\section{Noncommutative geometry and smeared matter distributions}
\label{sec:noncomm}

Noncommutative geometry provides a conceptually different route from
the generalized uncertainty principle discussed in the previous
section. Rather than modifying the phase-space structure of matter, the
basic idea is that spacetime coordinates themselves need not commute at
sufficiently short distances. A commonly used schematic relation is
\begin{equation}
[\hat{x}^{\mu},\hat{x}^{\nu}]
=
i\theta^{\mu\nu},
\label{eq:nc}
\end{equation}
where $\theta^{\mu\nu}$ is an antisymmetric tensor characterizing the
noncommutative structure~\cite{Nicolini2009,Hossenfelder2013}. Its
components have dimensions of length squared and therefore introduce
characteristic noncommutative scales of order
$\sqrt{|\theta^{\mu\nu}|}$.

Equation~\eqref{eq:nc} should be understood as a spacetime relation.
A complete noncommutative field theory generally requires a deformation
of the product of fields, for example through a star product, and its
phenomenology depends on the particular realization
~\cite{Nicolini2009}. In applications to static compact objects,
however, one often adopts a simpler effective description in which the
loss of perfect localization is represented by a spatial smearing of the
matter source. The discussion below refers to this
noncommutative-inspired construction and should not be interpreted as a
complete four-dimensional noncommutative theory of neutron-star matter.

The simplest example is a pointlike source. In a frequently used
prescription, the three-dimensional Dirac distribution is replaced by a
Gaussian kernel,
\begin{equation}
\delta^{(3)}(\mathbf{r})
\longrightarrow
K_{\theta}(r)
=
\frac{1}{(4\pi\theta)^{3/2}}
\exp\!\left(
-\frac{r^2}{4\theta}
\right),
\label{eq:gaussian_kernel}
\end{equation}
where $\theta>0$ sets the effective smearing scale. A point mass $M$
is then represented by
\begin{equation}
\rho_{\theta}(r)
=
M K_{\theta}(r)
=
\frac{M}{(4\pi\theta)^{3/2}}
\exp\!\left(
-\frac{r^2}{4\theta}
\right).
\label{eq:gaussian}
\end{equation}
The characteristic width of the source is therefore of order
$\sqrt{\theta}$. In the formal limit
\begin{equation}
\theta\rightarrow0,
\label{eq:nc_commutative_limit}
\end{equation}
the Gaussian kernel approaches the ordinary Dirac distribution and the
point-source description is recovered.

For a neutron star, however, the situation is more subtle. The stellar
interior is already described by an extended energy-density profile
$\epsilon(r)$ rather than by a point source. Consequently,
Eq.~\eqref{eq:gaussian} cannot by itself be taken as a unique
noncommutative prediction for the matter distribution of a star.

A natural phenomenological extension is to smear the original energy
density according to
\begin{equation}
\epsilon_{\theta}(\mathbf{r})
=
\int d^3r'\,
K_{\theta}
\!\left(
|\mathbf{r}-\mathbf{r}'|
\right)
\epsilon(\mathbf{r}'),
\label{eq:extended_smearing}
\end{equation}
where $K_{\theta}$ is the Gaussian kernel introduced in
Eq.~\eqref{eq:gaussian_kernel}. This form makes the nonlocal character
of the prescription explicit: the effective density at a given point
receives contributions from a finite surrounding region whose size is
controlled by $\sqrt{\theta}$.

Equation~\eqref{eq:extended_smearing} is useful as a coordinate-space
model, but it is not by itself a covariant prescription for a
relativistic star. In a fully geometric treatment, both the integration
measure and the distance entering the kernel should be defined with
respect to the spatial geometry of the stellar configuration. The
expression above should therefore be regarded as a transparent
phenomenological realization of nonlocal smearing rather than as a
unique covariant consequence of Eq.~\eqref{eq:nc}.

For a static and spherically symmetric configuration, one may define an
effective enclosed mass from the smeared density,
\begin{equation}
m_{\theta}(r)
=
4\pi
\int_0^r
\frac{\epsilon_{\theta}(r')}{c^2}
\,r'^2\,dr'.
\label{eq:massNC}
\end{equation}
This illustrates directly how a redistribution of the energy density can
modify the mass profile that enters the stellar structure equations.
However, the mass function alone is not sufficient to define a
self-consistent equilibrium configuration.

The full effective stress-energy tensor must satisfy the conservation
condition
\begin{equation}
\nabla_{\mu}T^{\mu\nu}_{\rm eff}=0.
\label{eq:nc_conservation}
\end{equation}
Moreover, the effective source generated by a noncommutative-inspired
construction need not retain the perfect-fluid form. In particular, the
radial and tangential pressures may differ,
\begin{equation}
P_r(r)\neq P_t(r),
\label{eq:nc_anisotropy}
\end{equation}
so that the stress-energy tensor becomes anisotropic.

This has an immediate consequence for the stellar equilibrium problem.
For an anisotropic fluid, the radial conservation equation contains an
additional contribution proportional to $P_t-P_r$, and the usual
perfect-fluid TOV equation must be replaced by its anisotropic form.
Therefore, smearing the density profile while leaving the pressure
sector unchanged is not sufficient. The energy density, pressure
components, and conservation equations must be constructed
consistently within the same effective model.

This point is particularly important because the Gaussian smearing
prescription is not unique. Different noncommutative-inspired
constructions can generate different effective density profiles,
pressure anisotropies, and therefore different stellar structure
equations. The parameter $\theta$ should consequently be understood as
a model-dependent effective scale rather than as a universal
noncommutative parameter whose astrophysical interpretation is
independent of the underlying construction.

The scale hierarchy again determines whether the effect can become
observable. If the noncommutative length is genuinely of order the
Planck length,
\begin{equation}
\sqrt{\theta}
\sim
\ell_{\rm Pl},
\label{eq:nc_planck_scale}
\end{equation}
the corresponding modification is expected to be negligible on
neutron-star length scales. Observable changes in the stellar structure
would instead require an effective smearing scale much larger than
$\ell_{\rm Pl}$, or some additional mechanism capable of enhancing the
low-energy manifestation of the noncommutative physics. Bounds derived
from neutron stars should therefore be interpreted as constraints on an
effective noncommutative-inspired model rather than as direct
measurements of microscopic spacetime noncommutativity
~\cite{Nicolini2009,Hossenfelder2013}.

It is also useful to separate the smearing prescription from the
microscopic equation of state. The local thermodynamic relation
\begin{equation}
P=P(\epsilon)
\label{eq:nc_eos}
\end{equation}
is determined by the underlying description of dense matter, including
nuclear interactions, particle composition, beta equilibrium, charge
neutrality, and many-body effects
~\cite{Oertel2017,Baym2018,Burgio2021}. In the class of models discussed
here, noncommutative-inspired effects act instead through the effective
spatial distribution and stress-energy structure of the source.

A realistic neutron-star calculation must therefore combine two
ingredients consistently: a microscopic EOS for dense matter and an
effective stress-energy tensor implementing the chosen
noncommutative-inspired prescription. Only after both are specified can
one solve the corresponding stellar structure equations and determine
observable quantities such as the mass, radius, or tidal response.

The comparison with the GUP case highlights the usefulness of the
classification introduced in Sec.~\ref{sec:framework}. GUP-inspired
models can modify the microscopic thermodynamics and hence the EOS,
whereas the noncommutative-inspired construction considered here acts
primarily through the localization and stress-energy structure of the
source. The next section considers the third route, in which the
gravitational dynamics themselves are modified through higher-curvature
terms and other effective extensions of GR.

\section{Effective field theory and modified gravity}
\label{sec:eft}

The previous sections considered corrections that enter through the
microscopic matter sector or through the effective stress-energy source.
A third possibility is that the gravitational dynamics themselves are
modified. At energies well below the scale at which the underlying
high-energy theory becomes relevant, such effects can be organized
systematically within an effective field theory (EFT) description of
gravity~\cite{Donoghue1994,Burgess2004}.

The EFT approach does not require a complete knowledge of the
ultraviolet theory. Instead, one writes the most general low-energy
action compatible with the symmetries of the theory and organizes the
allowed operators according to their number of derivatives, or
equivalently their mass dimension. For gravity, general covariance
allows scalar combinations of curvature tensors and their covariant
derivatives. The Einstein--Hilbert term provides the leading
contribution at low curvature, while higher-curvature and
higher-derivative operators encode corrections associated with shorter
distance scales.

Throughout this section we use natural units,
\begin{equation}
\hbar=c=1.
\label{eq:natural_units}
\end{equation}

A convenient schematic form of the local gravitational EFT is
\begin{equation}
\begin{aligned}
S_{\rm EFT}
={}&
\frac{\bar M_{\rm Pl}^{\,2}}{2}
\int d^4x\,\sqrt{-g}\,
\Bigg[
R
+\frac{c_1}{M^2}R^2
+\frac{c_2}{M^2}R_{\mu\nu}R^{\mu\nu}
\\
&\hspace{2.0cm}
+\frac{d_1}{M^4}\mathcal{O}_6
+\frac{d_2}{M^6}\mathcal{O}_8
+\cdots
\Bigg]
+
S_{\rm matter},
\end{aligned}
\label{eq:eftaction}
\end{equation}
where
\begin{equation}
\bar M_{\rm Pl}
=
\frac{1}{\sqrt{8\pi G}}
\label{eq:reduced_planck_mass}
\end{equation}
is the reduced Planck mass. The scale $M$ characterizes the heavy
degrees of freedom that are not resolved explicitly in the low-energy
description, while $c_i$ and $d_i$ are dimensionless Wilson
coefficients. The quantities $\mathcal{O}_6$ and $\mathcal{O}_8$
represent local curvature invariants of mass dimension six and eight,
respectively. For example, $\mathcal{O}_6$ may contain terms cubic in
the curvature.

The dimensional hierarchy makes the derivative expansion transparent.
In natural units, the Ricci scalar has mass dimension two,
\begin{equation}
[R]=(\text{mass})^2,
\end{equation}
and therefore
\begin{equation}
[R^2]=(\text{mass})^4,
\qquad
[\mathcal{O}_6]=(\text{mass})^6,
\qquad
[\mathcal{O}_8]=(\text{mass})^8.
\end{equation}
The inverse powers of the heavy scale $M$ therefore suppress operators
containing additional powers of the curvature or additional
derivatives.

Equation~\eqref{eq:eftaction} is useful for power counting, but it
should not be interpreted as a unique or minimal operator basis. In
four spacetime dimensions, the Gauss--Bonnet identity relates the
quadratic curvature invariants, while field redefinitions can move
contributions between different operators. In the presence of matter,
such redefinitions may also relate curvature terms to interactions
involving the matter fields~\cite{Donoghue1994,Burgess2004}.

Varying the action with respect to the metric leads schematically to
\begin{equation}
G_{\mu\nu}
+
\Delta_{\mu\nu}
=
\frac{1}{\bar M_{\rm Pl}^{\,2}}T_{\mu\nu}
=
8\pi G\,T_{\mu\nu},
\label{eq:modified_einstein}
\end{equation}
where $\Delta_{\mu\nu}$ collects the contributions generated by the
higher-order operators. Its explicit form depends on the terms retained
in the effective action. In this class of models, the microscopic
equation of state may remain unchanged even though the relation between
the matter source and the spacetime geometry is modified.

A familiar example is
\begin{equation}
f(R)
=
R+\alpha R^2,
\label{eq:starobinsky_example}
\end{equation}
where $\alpha$ has dimensions of inverse mass squared, or equivalently
length squared. This expression provides a simple illustration of how a
curvature-squared contribution can modify the gravitational field
equations while preserving general covariance
~\cite{SotiriouFaraoni2010,DeFeliceTsujikawa2010}.

An important distinction must be made between using
Eq.~\eqref{eq:starobinsky_example} as part of a low-energy expansion
and treating it as a complete modified-gravity theory. In the EFT
interpretation, the $R^2$ term is a correction whose importance is
controlled by the ratio between the characteristic curvature and the
scale suppressing the operator. The same functional form can instead
be studied nonperturbatively as a particular theory of gravity. Here it
is used only to illustrate how corrections in the gravitational sector
can enter the neutron-star problem.

More general theories may contain additional curvature invariants,
covariant derivatives, extra propagating degrees of freedom, or
nonminimal couplings to matter. There is therefore no universal
``modified TOV equation.'' The appropriate stellar-structure equations
must be derived from the field equations and conservation laws of the
specific theory under consideration.

At the level of hydrostatic equilibrium, a departure from GR may be
represented schematically as
\begin{equation}
\frac{dP}{dr}
=
F_{\rm GR}
+
\delta F,
\label{eq:modified_tov_schematic}
\end{equation}
where $F_{\rm GR}$ denotes the standard GR hydrostatic-balance term and
$\delta F$ represents the correction generated by the modified field
equations. The quantity $\delta F$ is not an arbitrary additional
force: it must follow consistently from the underlying theory and its
associated conservation equations.

\begin{figure}[t]
\centering
\includegraphics[width=\columnwidth]
{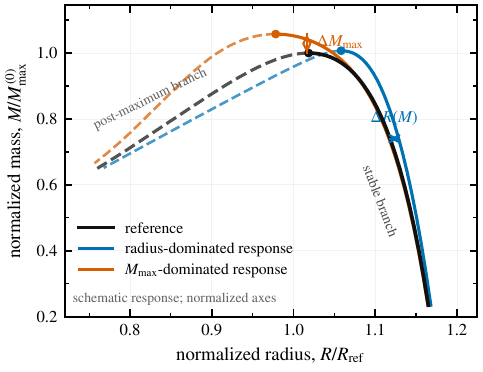}
\caption{
Schematic response of a neutron-star mass--radius sequence to changes
in the underlying stellar model. The black curve represents a
normalized reference sequence, while the additional curves illustrate
a predominantly radius-dominated response and a modification of the
maximum supported mass. Solid portions denote the stable branch and
dashed portions the post-maximum branch. The axes are normalized and
the curves are illustrative only; they are not solutions obtained from
a specific equation of state or quantum-gravity-inspired model.
}
\label{fig:mr_response}
\end{figure}

Changes in the gravitational dynamics can shift the mass--radius
relation, modify the maximum supported mass, and alter the compactness
and tidal response of the star~\cite{Berti2015,YagiYunes2013}.
Figure~\ref{fig:mr_response} illustrates two generic consequences that
are particularly relevant for observations: a displacement of the
radius at fixed mass and a change in the maximum mass.

\subsection{Power counting and the expected size of the corrections}
\label{subsec:eft_power_counting}

A central advantage of the EFT framework is that it provides a direct
estimate of the expected size of higher-order corrections. To
distinguish the characteristic curvature scale from the Ricci scalar
$R$ and from the stellar radius, we denote it by $\mathcal{R}$. In
natural units,
\begin{equation}
[\mathcal{R}]
=
(\text{mass})^2.
\label{eq:curvature_dimension}
\end{equation}
The Einstein--Hilbert contribution therefore scales as
\begin{equation}
\mathcal{L}_{\rm EH}
\sim
\bar M_{\rm Pl}^{\,2}\mathcal{R}.
\label{eq:eh_scaling}
\end{equation}

A curvature-squared term in Eq.~\eqref{eq:eftaction} scales
schematically as
\begin{equation}
\delta\mathcal{L}_{R^2}
\sim
\bar M_{\rm Pl}^{\,2}
\frac{c_i}{M^2}
\mathcal{R}^{\,2},
\label{eq:r2_scaling}
\end{equation}
so that, relative to the Einstein--Hilbert term,
\begin{equation}
\frac{\delta\mathcal{L}_{R^2}}
{\mathcal{L}_{\rm EH}}
\sim
c_i
\frac{\mathcal{R}}{M^2}.
\label{eq:eft_power_ratio}
\end{equation}
The dimensionless quantity
\begin{equation}
\frac{\mathcal{R}}{M^2}
\label{eq:eft_expansion_parameter}
\end{equation}
therefore controls the derivative expansion. When
$\mathcal{R}\ll M^2$, curvature-squared corrections remain
perturbative unless the associated Wilson coefficient is
parametrically large.

The same reasoning applies to higher powers of the curvature. A local
operator containing $n$ powers of the curvature scales schematically as
\begin{equation}
\delta\mathcal{L}_{R^n}
\sim
\bar M_{\rm Pl}^{\,2}
\frac{c_n}{M^{2n-2}}
\mathcal{R}^{\,n},
\qquad
n\geq2,
\label{eq:eft_general_operator}
\end{equation}
which gives
\begin{equation}
\frac{\delta\mathcal{L}_{R^n}}
{\mathcal{L}_{\rm EH}}
\sim
c_n
\left(
\frac{\mathcal{R}}{M^2}
\right)^{n-1}.
\label{eq:eft_general_power}
\end{equation}
For instance, a curvature-cubed contribution is suppressed by two
powers of the same expansion parameter,
\begin{equation}
\frac{\delta\mathcal{L}_{R^3}}
{\mathcal{L}_{\rm EH}}
\sim
c_3
\left(
\frac{\mathcal{R}}{M^2}
\right)^2.
\label{eq:eft_r3_power}
\end{equation}
Higher-curvature terms are therefore progressively suppressed as long
as the characteristic curvature remains well below the scale $M^2$.

If the scale associated with new physics is of order the reduced
Planck mass,
\begin{equation}
M\sim\bar M_{\rm Pl},
\end{equation}
the leading curvature-squared correction becomes
\begin{equation}
\frac{\delta\mathcal{L}_{R^2}}
{\mathcal{L}_{\rm EH}}
\sim
c_i
\frac{\mathcal{R}}
{\bar M_{\rm Pl}^{\,2}}.
\label{eq:eft_planck_suppression}
\end{equation}
For neutron-star spacetimes,
\begin{equation}
\frac{\mathcal{R}}
{\bar M_{\rm Pl}^{\,2}}
\ll 1,
\label{eq:ns_curvature_planck_ratio}
\end{equation}
so corrections suppressed directly by the Planck scale are expected to
be extremely small when the Wilson coefficients are of natural size.

This is the gravitational analogue of the scale hierarchy encountered
in the GUP and noncommutative-inspired cases. Neutron stars are strongly
relativistic on astrophysical scales, but their characteristic
curvatures remain far below the Planck scale. Observable deviations
therefore require an additional ingredient, such as a lower effective
scale $M$, enhanced Wilson coefficients, additional light degrees of
freedom, or another mechanism that amplifies the low-energy response.

Truncating a higher-derivative action and then solving it
nonperturbatively can introduce solution branches outside the EFT.
A controlled treatment retains solutions continuously connected to GR,
commonly through order reduction or an equivalent expansion. This
prevents degrees of freedom created by the truncation from being
mistaken for low-energy predictions \cite{Cayuso2023}.

The local curvature expansion is not the complete gravitational EFT.
Operators containing covariant derivatives of the curvature appear at
higher orders, while quantum loops generate nonlocal contributions that
cannot be represented by a finite polynomial of local curvature
invariants~\cite{Donoghue1994,Burgess2004}. The discussion above is
therefore intended only to describe the power counting of the local
derivative expansion relevant to the present analysis.

\subsection{Scale-dependent gravitational couplings}
\label{subsec:running_g}

A conceptually different possibility arises in renormalization-group
approaches to gravity, where Newton's coupling depends on a
renormalization scale,
\begin{equation}
G
\longrightarrow
G(k),
\label{eq:running_G}
\end{equation}
with $k$ denoting the RG scale
~\cite{Reuter1998,BonannoReuter2000}. This should be distinguished from
the local EFT expansion in Eq.~\eqref{eq:eftaction}. The latter
organizes higher-derivative operators in the effective action, whereas
$G(k)$ describes the scale dependence of a coupling under
renormalization-group evolution.

Applying a running gravitational coupling to a neutron star requires an
additional physical prescription because the RG scale $k$ must be
related to a characteristic scale of the stellar configuration. Possible
choices include a local curvature invariant, a density scale, or an
inverse characteristic length. There is no unique identification, and
different scale-setting prescriptions can lead to different stellar
solutions.

For the same reason, it is generally not sufficient to replace $G$ by a
position-dependent function directly in the standard Einstein
equations. The resulting field equations must remain compatible with
the Bianchi identities and with the appropriate conservation equations
for matter and the effective gravitational sector. A consistent stellar
calculation therefore requires three ingredients: the running law
$G(k)$, a prescription relating $k$ to the stellar configuration, and
the corresponding modified field equations.

Once these ingredients are specified consistently, the resulting
equilibrium solutions can be confronted with neutron-star observables
such as masses, radii, compactnesses, and tidal deformabilities
~\cite{Berti2015,YagiYunes2013}. As in the previous examples, the
resulting bounds apply to the particular low-energy realization being
tested and should not automatically be interpreted as direct
measurements of Planck-scale quantum gravity.

\section{Observational channels and prospective constraints}
\label{sec:observations}

The phenomenological relevance of the quantum-gravity-inspired
scenarios discussed in the previous sections ultimately depends on
whether their effects can be distinguished from uncertainties in
conventional neutron-star physics. This requires connecting the modified
microscopic or gravitational description to quantities that can be
measured astrophysically. The most useful information currently comes
from three complementary channels: precision measurements of
neutron-star masses, X-ray constraints on masses and radii, and
gravitational-wave measurements of the tidal response of neutron stars
in compact binaries.

For a specified microscopic EOS, effective source, and theory of
gravity, the stellar structure equations determine a family of
equilibrium configurations. From this sequence one obtains quantities
such as the mass $M$, radius $R$, compactness $C$, maximum supported
mass $M_{\rm max}$, and tidal deformability $\Lambda$. A
quantum-gravity-inspired parameter therefore enters the comparison with
observations only indirectly. Schematically,
\begin{equation}
\alpha_{\rm QG}
\longrightarrow
\left\{
P(\epsilon),\,
T_{\mu\nu}^{\rm eff},\,
\text{gravity}
\right\}
\longrightarrow
\left\{
M,\,
R,\,
\Lambda,\ldots
\right\}.
\label{eq:qg_to_observables}
\end{equation}
If the deformation acts only on the matter sector, the modified EOS is
inserted into the standard GR stellar equations. If the gravitational
sector is changed, both the equilibrium equations and, where relevant,
their perturbations must be derived and solved within the same theory.
Likewise, an effective modification of the matter source must be
implemented consistently in the stress-energy tensor entering the
stellar problem.

\subsection{Mass constraints}
\label{subsec:mass_constraints}

Neutron-star masses provide one of the most direct constraints on
stellar models because several pulsar masses are known with high
precision. The existence of neutron stars with masses close to or above
$2\,M_\odot$ is now firmly established
~\cite{Demorest2010,Antoniadis2013,Cromartie2020,Fonseca2021}.
Representative examples include
\begin{align}
M_{\rm J1614-2230}
&\simeq
1.97\,M_\odot,
\\
M_{\rm J0348+0432}
&\simeq
2.01\,M_\odot,
\\
M_{\rm J0740+6620}
&\simeq
2.08\,M_\odot.
\label{eq:two_solar_mass_pulsars}
\end{align}
These measurements imply the robust requirement
\begin{equation}
M_{\rm max}
\gtrsim
2\,M_\odot
\label{eq:mmax_constraint}
\end{equation}
for any viable combination of dense-matter physics and gravitational
dynamics.

Additional information comes from very massive spider pulsars.
PSR J0952$-$0607 has been reported with a mass around
$2.35\,M_\odot$, although its inference is more model dependent than
the Shapiro-delay measurements of pulsar--white-dwarf systems
~\cite{Romani2022}. Such systems provide potentially stronger
constraints on the upper end of the neutron-star mass distribution,
but it is useful to distinguish them from the more robust
$\sim2\,M_\odot$ lower bound.

A quantum-gravity-inspired correction that strongly softens the
effective EOS or increases the gravitational attraction sufficiently to
lower $M_{\rm max}$ below the observed range is therefore disfavored.
The interpretation is nevertheless not unique. A modification of
gravity that lowers $M_{\rm max}$ can, in principle, be compensated by
a sufficiently stiff microscopic EOS, while a change in the matter
sector can mimic the effect of altered gravitational dynamics. Mass
measurements are therefore powerful consistency tests, but they cannot
by themselves identify the microscopic origin of a deviation.

\subsection{Radius measurements}
\label{subsec:radius_constraints}

Radius measurements provide complementary information because they
probe the pressure of dense matter over a different range of densities
from that controlling the maximum mass. X-ray pulse-profile modeling
with the Neutron Star Interior Composition Explorer (NICER) has provided
simultaneous mass--radius constraints for several millisecond pulsars.
The first major results were obtained for PSR J0030+0451 and the massive
pulsar PSR J0740+6620
~\cite{Miller2019,Riley2019,Miller2021,Riley2021}.

The observational picture has continued to develop. Updated analyses of
PSR J0740+6620 using substantially larger NICER data sets have refined
its radius constraint~\cite{Salmi2024,Dittmann2024}, while NICER
observations of the nearby pulsar PSR J0437$-$4715 have provided an
independent radius measurement at a mass close to
$1.4\,M_\odot$~\cite{Choudhury2024}. These newer results broaden the
mass range over which the neutron-star mass--radius relation can be
tested directly.

Taken together with nuclear and gravitational-wave information, current
analyses typically favor radii near $\sim12$ km for canonical-mass
neutron stars, with uncertainties at the kilometer level and some
dependence on the EOS parametrization and adopted data sets
~\cite{Raithel2019,Annala2020,Rutherford2024}. It is therefore more
appropriate to retain a broad characteristic scale,
\begin{equation}
R_{\rm NS}
\sim
11\text{--}14\ {\rm km},
\label{eq:radius_scale}
\end{equation}
rather than associate neutron-star phenomenology with a single
preferred radius.

For the present discussion, the important point is how a new physical
ingredient shifts the mass--radius relation. A change in the microscopic
EOS can modify the stellar radius at fixed mass, while a modification
of the gravitational field equations can produce a similar displacement
even for the same EOS. The same is true for changes in the effective
stress-energy source. Radius information is therefore most useful when
combined with mass measurements and with observables that probe the
stellar response in a different way.

\subsection{Tidal deformability}
\label{subsec:tidal_constraints}

Gravitational waves provide such a complementary probe through the tidal
response of neutron stars in binary systems. During the inspiral, the
external tidal field generated by one star induces a multipolar
deformation in its companion. In GR, the dimensionless quadrupolar tidal
deformability is
\begin{equation}
\Lambda
=
\frac{2}{3}
k_2
\left(
\frac{Rc^2}{GM}
\right)^5
=
\frac{2}{3}
k_2 C^{-5},
\label{eq:tidal_deformability}
\end{equation}
where $k_2$ is the quadrupolar tidal Love number and
\begin{equation}
C
=
\frac{GM}{Rc^2}
\label{eq:tidal_compactness}
\end{equation}
is the stellar compactness
~\cite{Berti2015,YagiYunes2013,Chatziioannou2020}.

\begin{figure}[t]
\centering
\includegraphics[width=\columnwidth]
{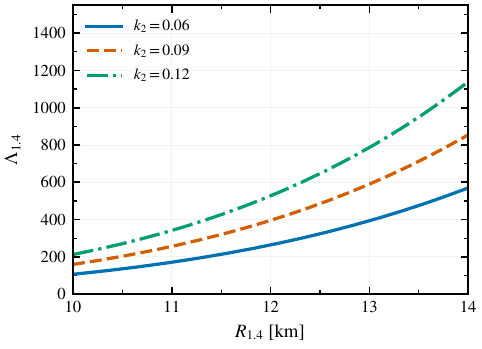}
\caption{
Dimensionless tidal deformability $\Lambda_{1.4}$ as a function of the
radius $R_{1.4}$ for a $1.4\,M_\odot$ neutron star and representative
fixed values of the quadrupolar Love number $k_2$. In GR,
$\Lambda=(2/3)k_2C^{-5}$, so the tidal response is strongly amplified by
changes in stellar compactness. The curves are intended to illustrate
this sensitivity and are not predictions of a particular EOS or
quantum-gravity-inspired model.
}
\label{fig:tidal_response}
\end{figure}

The strong $C^{-5}$ dependence makes $\Lambda$ particularly sensitive
to the stellar radius. Even a moderate change in compactness can
therefore produce a substantial change in the tidal response. This
sensitivity is illustrated in Fig.~\ref{fig:tidal_response} for fixed
representative values of $k_2$.

The binary neutron-star merger GW170817 provided the first
gravitational-wave measurement with clear sensitivity to neutron-star
tidal effects~\cite{Abbott2017,Abbott2018}. The subsequent detection of
GW190425 enlarged the binary-neutron-star sample, although its tidal
information was considerably less constraining
~\cite{Abbott2020GW190425}. Joint analyses of gravitational-wave,
NICER, pulsar-mass, and nuclear-theory information now provide stronger
constraints on the dense-matter EOS than any of these channels alone
~\cite{Pang2021,Rutherford2024}.

A commonly quoted reference quantity is the tidal deformability of a
$1.4\,M_\odot$ neutron star. For the purposes of the present discussion,
the relevant scale remains
\begin{equation}
\Lambda_{1.4}
\sim
\mathcal{O}(10^2\text{--}10^3),
\label{eq:lambda_scale}
\end{equation}
while the precise credible interval depends on the waveform model,
priors, and assumptions used to parametrize the EOS.

An additional subtlety arises when the gravitational sector is modified.
Equation~\eqref{eq:tidal_deformability} may still provide a useful
definition of a dimensionless tidal response, but the value of $k_2$
cannot in general be taken from the corresponding GR solution.
Modified gravity can change both the equilibrium background and the
linear perturbation equations from which the Love number is obtained.
A consistent comparison with gravitational-wave data therefore requires
the stellar background and tidal perturbations to be calculated within
the same gravitational framework.

\subsection{Combining the observables}
\label{subsec:multimessenger_combination}

The main strength of neutron-star phenomenology comes from combining
observables that respond differently to the microscopic EOS and the
gravitational dynamics. Massive pulsars constrain the maximum mass,
X-ray observations restrict the mass--radius relation, and
gravitational-wave measurements probe the tidal response. Modern
multimessenger analyses combine these observations with theoretical
information from nuclear physics to infer the EOS over an increasingly
broad density range~\cite{Pang2021,Rutherford2024}.

A viable model must therefore reproduce several observables
simultaneously, which can be summarized schematically as
\begin{equation}
\left\{
M_{\rm max},
R(M),
\Lambda(M)
\right\}_{\rm theory}
\sim
\left\{
M,
R,
\Lambda
\right\}_{\rm obs},
\label{eq:multimessenger_constraint}
\end{equation}
where the symbol $\sim$ denotes consistency within the corresponding
observational and theoretical uncertainties.

The need for a combined analysis follows from a central degeneracy in
neutron-star physics. Changes in nuclear interactions, many-body
correlations, particle composition, or high-density phase structure can
shift $M$, $R$, and $\Lambda$ in ways that resemble the effects of
modified gravity or other new physics
~\cite{Oertel2017,Baym2018,Burgio2021,Raithel2019}. A displaced
mass--radius curve, for example, does not by itself reveal whether the
origin of the shift lies in the EOS or in the gravitational sector.

This issue is particularly relevant for the three classes of models
considered here. In the GUP realization of Sec.~\ref{sec:gup}, the
correction enters through the microscopic thermodynamics and therefore
changes the EOS. In the noncommutative-inspired construction of
Sec.~\ref{sec:noncomm}, the effective spatial distribution and
stress-energy structure of the source are modified. In the
gravitational EFT and related scenarios of Sec.~\ref{sec:eft}, the
field equations and stellar equilibrium conditions may change directly.
Despite their different microscopic interpretations, all three
mechanisms ultimately feed into the same limited set of macroscopic
observables.

Multimessenger observations should therefore not be expected to reveal a
unique quantum-gravity signature in isolation. Their more realistic role
is to reduce the region of parameter space in which a given
quantum-gravity-inspired scenario remains compatible with neutron-star
data. A meaningful constraint must be obtained together with the
uncertainties of the dense-matter EOS and, when the gravitational sector
is modified, with a self-consistent treatment of both the equilibrium
configuration and its perturbations.

As measurements of neutron-star masses, radii, and tidal properties
improve, this combined approach will become increasingly restrictive.
Its value lies not in providing direct access to Planck-scale physics,
but in testing how large an effective departure from the standard
description of dense matter and gravity can remain compatible with the
observed properties of neutron stars.

\subsection{A minimum standard for a credible bound}
\label{subsec:credible_bound}
The EOS--gravity degeneracy is not a small correction: variations of the
high-density symmetry energy can exceed changes produced by some
allowed alternative-gravity models \cite{HeFattoyevLiNewton2015}.
A credible bound should state (i) the algebra, action, or source being
constrained; (ii) the controlled expansion parameter; (iii) the
thermodynamic and conservation equations; (iv) the EOS and composition
range; (v) whether background and tides use the same theory; and
(vi) the likelihoods and priors. Without these items, a limit may
constrain a toy model but cannot be transferred to another GUP algebra,
smearing prescription, or EFT coefficient.

\section{Conclusions}
\label{sec:conclusions}

Neutron stars occupy an unusual position in the search for physics
beyond the standard descriptions of matter and gravity. They combine
supranuclear densities, strong gravitational fields, and a growing set
of precision observations, yet they remain enormously separated from
the Planck regime. Their relevance to quantum-gravity phenomenology is
therefore not that they probe the microscopic structure of spacetime
directly, but that they provide a controlled astrophysical environment
in which possible low-energy remnants of such physics can be tested.

The examples discussed in this article illustrate three distinct routes
through which quantum-gravity-inspired effects can enter the
neutron-star problem. Generalized uncertainty principles can modify the
microscopic phase-space structure and consequently the thermodynamics
and equation of state. Noncommutative-inspired constructions provide a
different possibility, in which the effective localization and
stress-energy distribution of the matter source are altered.
Higher-curvature corrections and scale-dependent gravitational
couplings instead modify the gravitational dynamics and therefore the
equations governing stellar equilibrium. This separation into
matter-sector, effective-source, and gravitational corrections provides
a useful framework for comparing models whose microscopic motivations
may otherwise be very different.

A common conclusion emerges from all three cases. If the relevant
corrections are genuinely controlled by the Planck scale and their
dimensionless coefficients are of natural size, their effects under
neutron-star conditions are expected to be extremely small. Observable
changes in stellar structure generally require an additional ingredient:
a lower effective scale, enhanced couplings, new low-energy degrees of
freedom, or some other mechanism that amplifies the correction far below
the Planck regime. Astrophysical bounds should therefore be interpreted
primarily as constraints on the effective parameters of a particular
phenomenological realization, rather than as direct measurements of a
fundamental quantum-gravity scale.

The interpretation of such bounds is further complicated by the
uncertainty of the high-density equation of state. Changes in nuclear
interactions, particle composition, many-body correlations, or possible
phase transitions can modify the mass--radius relation and tidal
response in ways that resemble the effects of new gravitational or
microscopic physics. No single neutron-star observable can therefore be
expected to provide an unambiguous quantum-gravity signature. Reliable
constraints require the dense-matter model, the effective source, and
the gravitational dynamics to be treated consistently within the same
stellar calculation.

This is where multimessenger observations become particularly
important. Precise pulsar masses constrain the maximum mass that a
viable model must support, X-ray observations probe the mass--radius
relation, and gravitational-wave measurements provide complementary
information through the tidal response. Their combined use can reduce
degeneracies that remain unavoidable when any one observable is
considered in isolation. The same consistency requirement applies on
the theoretical side: when gravity is modified, both the equilibrium
configuration and the corresponding perturbation equations must be
derived within the same framework.

The most promising role of neutron stars in quantum-gravity
phenomenology is therefore one of exclusion and consistency. Rather
than providing direct access to Planck-scale physics, they can determine
how large an effective departure from conventional dense-matter physics
and General Relativity can remain compatible with observation. As
mass and radius measurements improve, gravitational-wave observations
accumulate, and the high-density equation of state becomes better
constrained, this allowed parameter space will continue to shrink.

Future progress will depend as much on theoretical consistency as on
observational precision. Meaningful tests require models in which the
connection between the microscopic deformation, its effective
low-energy scale, the stellar structure equations, and the predicted
observables is explicit. Establishing this connection is essential if
neutron-star constraints are to be interpreted not merely as bounds on
phenomenological parameters, but as useful information about the
possible low-energy structure of a more fundamental theory of gravity.

\section*{Funding}
This research was funded by the Coordena\c{c}\~ao de Aperfei\c{c}oamento de Pessoal de N\'ivel Superior (CAPES), Brazil, Grant No.~88887.893649/2023-00.

\section*{Data availability}
No research datasets were generated or analyzed in this study. The
numerical data and plotting scripts used for the illustrative figures
are available from the author upon reasonable request.

\section*{Acknowledgements}
The author thanks colleagues and collaborators for helpful discussions
and comments. 
\section*{Conflict of interest}
The author declares no conflicts of interest.

\end{document}